\documentclass[nofootinbib,superscriptaddress,twocolumn]{revtex4-1}

\usepackage{amsmath}
\usepackage{amsfonts}
\usepackage{times}
\usepackage{graphicx}

\begin{document}

\title{Weakness of accelerator bounds
 on electron superluminality without a preferred frame}

\author{Giovanni AMELINO-CAMELIA}
\affiliation{Dipartimento di Fisica, Universit\`a di Roma ``La Sapienza", P.le A. Moro 2, 00185 Roma, EU}
\affiliation{INFN, Sez.~Roma1, P.le A. Moro 2, 00185 Roma, EU}

\author{Giulia GUBITOSI}
\affiliation{Department of Physics, University of California, Berkeley,
CA 94720, USA}
\affiliation{PCCP and APC, Universit/'e Denis Diderot Paris 7, 
Batiment Condorcet, 10 rue Alice Domon et L\'eonie Duquet, 75205 Paris, France}

\author{Niccol\'o~LORET}
\affiliation{Dipartimento di Fisica, Universit\`a di Roma ``La Sapienza", P.le A. Moro 2, 00185 Roma, EU}
\affiliation{INFN, Sez.~Roma1, P.le A. Moro 2, 00185 Roma, EU}

\author{Flavio MERCATI}
\affiliation{Departamento de F\'isica T\'eorica, Universidad de Zaragoza, Zaragoza 50009, Spain}

\author{Giacomo ROSATI}
\affiliation{Dipartimento di Fisica, Universit\`a di Roma ``La Sapienza", P.le A. Moro 2, 00185 Roma, EU}
\affiliation{INFN, Sez.~Roma1, P.le A. Moro 2, 00185 Roma, EU}

\begin{abstract}
\noindent
The reference laboratory bounds on superluminality of the electron
are obtained from the absence of in-vacuo Cherenkov processes
and the determinations of synchrotron radiated power for LEP electrons.
It is usually assumed that these analyses establish the validity of a standard special-relativistic
description of the electron with accuracy of at least a few parts in $10^{14}$,
and in particular this is used to exclude electron superluminality with such an accuracy.
We observe that these bounds rely crucially on the availability of a preferred frame.
In-vacuo-Cherenkov processes are automatically forbidden in any theory
with ``deformed Lorentz symmetry", relativistic theories that, while different from Special Relativity, preserve the relativity of inertial frames.
Determinations  of the synchrotron radiated power can be used to constrain
the possibility of Lorentz-symmetry deformation,
but provide rather weak bounds, which in particular for electron
superluminality we establish to afford us no more constraining power than for an accuracy
of a few parts in $10^4$. We argue that this observation can have only a limited role
in the ongoing effort of analysis of the anomaly tentatively reported by the OPERA collaboration,
but we stress
that it could provide a valuable case study for
assessing the limitations of ``indirect" tests
of fundamental laws of physics.
\end{abstract}

\maketitle

\baselineskip11pt plus .5pt minus .5pt

\section{Introduction and motivation}
Arguably the most classic task of physics research are tests
of the laws we adopt (at a given point in the history of physics)
as ``fundamental". This is where the dividing line between science and
other human endeavors is more vividly manifest: the ``truths" of science
are stubbornly scrutinized; ``true" in science only means not yet established
to be false. Of course, this does not affect the robustness of what our efforts
produce: our laws remain valid in the regimes used to establish them
({\it e.g.}, Galilean relativity still governs unchallenged the classes of
measurements and measurement accuracies available to Galileo and contemporaries),
but it characterizes the prudent undogmatic attitude we adopt when exploring
new realms of physics.

The analysis we here report is relevant for this defining aspect of fundamental physics,
and specifically for the
difference between direct and indirect tests of fundamental laws. The pretext
for us to intervene on these themes is provided by the
Lorentz-symmetry anomaly that was recently tentatively reported by
the OPERA collaboration~\cite{opera}.
Chances are the OPERA anomaly is going to be addressed
by far less exotic descriptions than departures
from Lorentz symmetry (such as aspects of the analysis
of systematic uncertainties), but these days of analysis
of the OPERA anomaly are providing a very clear case study for
the usefulness of our determined efforts of testing ``fundamental"
laws, and also for the limitations to this usefulness
that result from relying too heavily on indirect tests.
For clarity and simplicity we focus here on one of the relevant
points: the ``amount of superluminality"
of neutrinos apparently reported by OPERA
far exceeds the bounds on superluminality of the electron that are most
frequently quoted, which are based on electron
studies at LEP~\cite{altschul,lehnert}. Among the many ways in which
the OPERA anomaly is ``paradoxical" it is typical to include also its
being an awkward match for the tight bounds on LEP-electron superluminality.
But these tight bounds have emerged from indirect tests of electron
superluminality, and of course indirect tests are in general strongly model dependent.
If the OPERA anomaly eventually proved to be sound it would certainly produce a formidable shock wave
in fundamental physics,
also reshaping our criteria for what may constitute ``plausible model building".
In light of this it is legitimate to question whether
 presently-available indirect (and therefore model-dependent)
bounds on superluminality can provide any useful guidance
for the assessment of the OPERA anomaly.

Indirect bounds on new effects may be somewhat more reliable in other areas
of physics, where the ``rules of the game" are not being questioned and one is only
speculating about possible additional ingredients that could be introduced. But bounds on violations
of fundamental laws are clearly different: the hypothesis being tested in such cases
is that one of the pillars of our present conceptualization of Nature might have
to be revised, and one cannot meaningfully test this sort of hypothesis
 while relying on the strategies for model building that are currently in fashion.
  
We are here reporting some observations that give
 tangibility to our concerns, taking as starting point our
 previous works on test theories for Lorentz symmetry
which do not introduce a preferred frame (see, {\it e.g.}, Refs.~\cite{dsr1,dsr2,dsrnature,bob}).
We intend to establish that the most cited information against superluminality
from accelerator-electron data actually crucially uses in the data analysis
the availability of a preferred frame. So these ``indirect" bounds are not bounds
on the validity of the second special-relativity postulate (the ``maximum speed postulate")
on its own, but rather primarily test the validity of the first postulate
(the ``relativity-principle postulate") and have some implications for superluminality
of the electron only conditionally to the assumption of a concurring violation
of the first postulate.

We argue that the thesis here elaborated should encourage a more determined effort of
direct tests of the special-relativistic speed law, undeterred by the (limited significance of)
apparently powerful indirect tests of this speed law. And we hope that this attitude
will be adopted much beyond the probably limited ``shelf time" of the OPERA anomaly.
The possibility that the superluminality interpretation of the OPERA data
ends up proving correct is very exciting, but we are here
assuming that this is very unlikely. And yet we view
the OPERA anomaly as an exciting opportunity, a sort of conceptual laboratory, for
a tune up of the different competing attitudes adopted in fundamental-physics research. We feel
 in particular that the attitude here adopted in contemplating the tentative OPERA
anomaly is valuably complementary to the one adopted in the recent Letter
of Ref.~\cite{cohenGla}.

\section{Deformations of Lorentz symmetry}\label{dsrgeneral}
The notion of deformed Lorentz symmetry which we shall here adopt in quantifying
our concerns is the one
of the proposal ``DSR" (``doubly-special", or ``deformed-special", relativity),
first introduced in Ref.~\cite{dsr1} (accompanied by the more focused, but more limited,
analysis of Ref.~\cite{dsr2}).\\
This proposal was put forward
as a possible
description of {\underline{preliminary}}
theory results suggesting that there {\underline{might}} be violations
of some special-relativistic laws in certain approaches to the quantum-gravity problem,
most notably the ones based on spacetime noncommutativity and loop quantum gravity.
The part of the quantum-gravity community interested in those results was interpreting them
as a manifestation of a full breakdown of Lorentz symmetry, with the emergence of
a preferred class of observers (an ``ether"). But it was argued in Ref.~\cite{dsr1}
that departures from Special Relativity governed by a high-energy/short-distance scale
may well be compatible with the Relativity Principle, the principle of relativity
of inertial observers, at the cost of allowing some consistent modifications
of the Poincar\'e transformations, and particularly of the Lorentz-boost transformations.

The main area of investigation of the DSR proposal has been for the last decade
the possibility of introducing relativistically some Planck-scale-deformed on-shell relations.
The  DSR proposal was put forward~\cite{dsr1} as a conceptual path for pursuing
a broader class of scenarios of interest for fundamental physics, with or without
quantum-gravity motivation, including the possibility of introducing the second
observer-independent scale primitively in spacetime structure or primitively at the
level of the (deformed) de Broglie relation between wavelength and momentum.
However, the bulk of the preliminary results providing encouragement for this
approach came from quantum-gravity research
concerning Planck-scale departures from the special-relativistic on-shell relation,
and this in turn became the main focus of DSR research.

This idea of deformed Lorentz symmetry is actually very simple, as we shall here render manifest
on the basis of an analogy with how the Poincar\'e transformations came to be adopted
as a deformation of Galileo transformations.
Famously, as the Maxwell formulation of electromagnetism,
with an observer-independent speed scale ``$c$", gained more and more
experimental support (among which we count the Michelson-Morley results)
it became clear that Galilean relativistic symmetries could no longer be upheld.
From a modern perspective we should see the pre-Einsteinian attempts to address that
crisis (such as the ones of Lorentz) as attempts to ``break Galilean invariance",
 {\it i.e.} preserve the validity of Galilean transformations
as laws of transformation among inertial observers, but renouncing to the possibility that those
transformations be a symmetry of the laws of physics. The ``ether" would be a preferred frame
for the description of the laws of physics, and the laws of physics that hold in other frames
would be obtained from the ones of the preferred frame via Galilean transformations.\\
Those attempts failed.
What succeeded is completely complementary. Experimental evidence, and the analyses
of Einstein (and Poincar\'e) led us to a ``deformation of Galilean invariance":
in Special Relativity the laws of transformation
among observers still are a symmetry of the laws of physics (Special Relativity is no less
relativistic then Galilean Relativity), but the special-relativistic transformation laws
are a $c$-deformation of the Galilean laws of transformation with the special property
of achieving the observer-independence of the speed scale $c$.

This famous $c$-deformation in particular replaces the Galilean on-shell relation
$E= \mathrm{constant} + {\bf p}^2/(2m)$ with the special-relativistic version
$$E= \sqrt{c^2{\bf p}^2+c^4 m^2}~,$$
and the Galilean
composition of velocities ${\bf u} \oplus {\bf v} = {\bf u} + {\bf v}$
with the special relatistic law of composition of velocities
\begin{equation}
\!\!\!\!\!\!\!\!\!\!\!\!\!\!
{\bf u} \oplus_c {\bf v} = \frac{1}{1+\frac{{\bf u} \cdot {\bf v}}{c^2}}
\left({\bf u} + \frac{1}{\gamma_u}  {\bf v}
+ \frac{1}{c^2} \frac{\gamma_u}{1+\gamma_u} ({\bf u} \cdot {\bf v}){\bf u}  \right)
\label{ungarVEL}
\end{equation}
where as usual $\gamma_u \equiv 1/\sqrt{1- {\bf u} \cdot {\bf u}/c^2}$.\\
The richness of the velocity-composition (\ref{ungarVEL})
is a necessary match for the demanding task of introducing an absolute scale in a relativistic theory.
And it is unfortunate that undergraduate textbooks often choose to limit the discussion to the
special case of (\ref{ungarVEL}) which applies when ${\bf u}$ and ${\bf v}$ are collinear:
\begin{equation}
{\bf u} \oplus_c {\bf v} \Big|_{collinear}
= \frac{{\bf u} + {\bf v}}{1+\frac{{\bf u} \cdot {\bf v}}{c^2}}~.
\label{velTextbook}
\end{equation}
The invariance of the velocity scale $c$ of course requires that boosts act non-linearly on
velocity space, and this is visible not only in (\ref{ungarVEL}) but also in (\ref{velTextbook}).
But also the non-commutativity and non-associativity
 of (\ref{ungarVEL}) (which are silenced in (\ref{velTextbook}))
play a central role~\cite{ungar,ungarFOLLOWER,florianeteraUNGAR}
 in the logical consistency of Special Relativity as a theory
enforcing relativistically the absoluteness of the speed scale $c$.
For example, the composition law (\ref{ungarVEL}) encodes Thomas-Wigner rotations,
and in turn the relativity of simultaneity.

Equipped with this quick reminder of some features of the transition from
Galilean Relativity to Special Relativity we can now quickly summarize
the logical ingredients of a DSR framework. The analogy is particularly close in cases
where the DSR-deformation of Lorentz symmetry is introduced primitively
at the level of the on-shell relation.
To see this let us consider an on-shell relation (from now on we adopt conventions
with $c=1$)
\begin{equation}
m^2 = p_0^2 - {\bf p}^2 + \Delta(E,{\bf p};M_*)
\label{dsr1gen}
\end{equation}
where $\Delta$ is the deformation and $M_*$ is the deformation scale.

Evidently when $\Delta \neq 0$ such an on-shell relation (\ref{dsr1gen})
is not Lorentz invariant. If we insist on this law and on
the validity of classical (undeformed) Lorentz transformations between inertial
observers we clearly end up with a preferred-frame picture, and the Principle
of Relativity of inertial frames must be abandoned: the scale $M_*$ cannot
be observer independent, and actually the whole form of (\ref{dsr1gen}) is subject
to vary from one class of inertial observers to another.\\
The other option~\cite{dsr1} in such cases is the DSR option of enforcing
the relativistic invariance of (\ref{dsr1gen}), preserving the relativity
of inertial frames, at the cost of modifying the action of boosts on momenta.
Then in such theories both the velocity scale $c$ (here mute only because of the
choice of dimensions) and the energy scale $M_*$ play the
same role~\cite{dsr1,dsrnature}
of invariant scales of the relativistic theory which govern the form of boost
transformations. \\
Several examples of boost deformations adapted in the DSR sense to modified on-shell
relations have been analyzed in some detail
(see {\it e.g.} Refs.~\cite{dsr1,dsr2,dsrnature,leeDSRprd,jurekDSR2,leeDSRrainbow,gacdsrrev2010}),
often borrowing mathematical structures known from
Hopf-algebra versions of the Poincar\'e (Lie-) algebra~\cite{majidruegg,kpoinap,lukie1992}.
Clearly these DSR-deformed boosts ${\cal N}_j$ must be such that
they admit the deformed shell, $p_0^2 - {\bf p}^2 + \Delta(E,{\bf p};M_*)$,
as an invariant,
and in turn the law of composition of momenta must also be
deformed~\cite{dsr1}, $p_\mu \oplus_{\cal N} k_\mu$,
since it must be covariant under the action of the (DSR-deformed) boost transformations.\\
All this is evidently analogous to corresponding aspects of Galilean Relativity and Special Relativity:
of course in all 3 cases the on-shell relation is boost invariant (but respectively under Galilean boosts,
Lorentz boosts, and DSR-deformed Lorentz boosts); for Special Relativity the action of boosts
evidently must depend on the speed scale $c$ and must be non-linear on velocities
(since it must enforce observer independence of $c$-dependent laws), and for DSR relativity
the action of boosts
evidently must depend on both the speed scale $c$ and the energy scale $M_*$, acting non-linearly
both on velocities and momenta, since it must enforce observer independence of $c$-dependent and $M_*$-dependent laws.

\section{No in-vacuo-Cherenkov processes}
A large number of the tests providing the basis of our
perceived trust in classical Lorentz symmetry is based
on analyses of ``anomalous thresholds"~\cite{gacPIRANprd2001}.
An anomalous threshold occurs when a full breakdown of Lorentz symmetry
(with emergence of a preferred ``ether" frame) produces the effect that
a process which is not allowed (respectively allowed) in Special Relativity
is still not allowed (resp allowed)
at sufficiently low energies  in the new Lorentz-breaking theory,
but above a certain threshold energy the process is instead allowed (resp not allowed).
This is the case of the ``in-vacuo-Cherenkov threshold" which
is credited~\cite{altschul,lehnert,colgla}
for providing a particularly stringent limit on superluminality of the electron
in scenarios with a preferred frame:
 above a certain threshold energy for the (presumedly)
 superluminal electron
the Cherenkov process $e^- \rightarrow e^- + \gamma$
is allowed for electrons propagating in vacuum.
The fact that such in-vacuo-Cherenkov processes (if at all present) did not produce
noticeable energy losses for LEP electrons  implies
that the threshold is higher than $\sim 100GeV$, which in turn allows us to conclude
that the parameters responsible for breaking Lorentz symmetry must be
correspondingly small~\cite{altschul,lehnert,colgla}.

These bounds on departures from Lorentz symmetry based on anomalous
thresholds are completely inapplicable to scenarios where the departures
from Lorentz symmetry do not produce a preferred-frame picture.
Threshold-energy requirements such as the in-vacuo-Cherenkov threshold
are of course not Lorentz invariant, unlike the (very different, indeed ``non-anomalous")
threshold-energy requirements that Special Relativity does predict for example
for processes such as $\gamma + \gamma \rightarrow  e^+  + e^-$.
And they evidently {\underline{require}} a preferred-frame formulation.
In order to see this more vividly one may consider the case of an
observer Alice for whom the electron propagates in vacuum with
energy just above the threshold
for in-vacuo Cherenkov, so Alice can observe Cherenkov radiation from that electron.
If this was not a preferred-frame picture there would then be
the paradox of
an observer Bob who would disagree on the presence of the Cherenkov radiation:
this would happen for any observer Bob boosted with respect to Alice 
in such a way that
according to Bob the relevant electron has
energy below the in-vacuo-Cherenkov
threshold.

We are therefore assured that there cannot be any anomalous thresholds in DSR theories,
since they do not admit a preferred-frame picture.
This was noticed already several years ago~\cite{gacnewjourn}, and has been
since also explicitly
verified in several
studies~\cite{gacnewjourn,sethmajor,operaDSR,goldenrule} by taking different
 examples of
mutually DSR-compatible combinations of deformed boosts $\cal N$ and deformed
laws of composition of momenta $\oplus_{\cal N}$, finding that indeed when this mutual
DSR-compatibility is enforced one has that the modifications of the on-shell relation
 combine with the corresponding modifications of the law of conservation
of total momentum (total momentum computed with $\oplus_{\cal N}$ composition law)
to produce no anomalous thresholds.\\
This point resurfaced recently
in the novel literature on superluminal neutrinos, where an anomalous
threshold for $\nu_\mu \rightarrow \nu_\mu  + e^+ + e^-$ (a ``Cherenkov-like"
process for neutrinos) could be relevant~\cite{cohenGla},
and it was once again noticed to be a concern that cannot apply
to DSR pictures~\cite{operaDSR}.

\section{Synchrotron radiated power in DSR}
In light of the results briefly reviewed in the previous section
it should be now clear that this Letter mainly is about
synchrotron radiation in scenarios with deformation of Lorentz symmetry,
which instead do in general provide bounds on such deformations.

For scenarios with a full breakdown of Lorentz symmetry, and therefore
a preferred frame, the determinations of synchrotron radiated power by LEP electrons
provide an ultra-stringent constraint on electron superluminality,
as emphasized recently by Altschul~\cite{altschul}. These determinations
agree with the corresponding special-relativistic prediction to better than 0.1\% accuracy.
More precisely the available data can be used to establish that~\cite{altschul}
\begin{equation}
\Big| \frac{P_{exp}-P_{SR}}{P_{SR}} \Big| < 6 \times 10^{-4}
\label{agreement}
\end{equation}
where of particular interest is the dependence of
the special-relativistic prediction $P_{SR}$
on the rapidity $\xi_{SR}$ which is needed to connect
the instantaneous rest frame of a LEP electron to its lab frame:
$P_{SR} = P_{0} \cosh^4 (\xi_{SR})$ (where $P_0$ is
obtained from computing the radiated power in the instantaneous
rest frame of a LEP electron).

One can then argue~\cite{altschul} that in the rest frame of the electron the
implications of electron superluminality, and therefore the departures
from special relativity, should be negligibly small, and this allows one
to focus exclusively on the factor $\cosh^4 (\xi_{SR})$ as the key for establishing
bounds: theories with departures from Lorentz symmetry would require a different
value of rapidity for connecting the lab frame to the instantaneous rest frame.
This is the line of analysis adopted by Altschul~\cite{altschul}, who properly specified
as reference framework a framework with a full breakdown of Lorentz symmetry, the
 so-called  Standard Model Extension (see, {\it e.g.},
Refs.~\cite{sme1,smerev}). For our purposes here it suffices to consider Altschul's
argument for the case of superluminal electrons within the Coleman-Glashow broken-Lorentz-symmetry
picture of Ref.~\cite{colgla}, which is the part of the Standard Model Extension
most intuitively connected
to the structure of the argument. This amounts to attributing to superluminal electrons
an on-shell relation
\begin{equation}
m^2 = E^2 - (1 + 2 \delta) {\bf p}^2~,
\label{colglaele}
\end{equation}
and then looking~\cite{altschul} for the rapidity $\xi_{LIV}$ that connects
such a superluminal LEP electron of 91 GeV to its rest frame, assuming (as standard for scenarios
with broken Lorentz symmetry and a preferred frame) that the boost transformations
are still governed by classical Lorentz boosts. Of course, classical Lorentz boosts
are a very awkward match for (\ref{colglaele}) (this is after all why a preferred
frame arises). As a result Altschul correctly finds~\cite{altschul} that the relationship
between $\xi_{LIV}$  and $\xi_{SR}$ manifests a large mismatch even for
small values of $\delta$:
\begin{equation}
\cosh^4 (\xi_{LIV}) \simeq \cosh^4 (\xi_{SR}) [1 + 4 \delta \cosh^2 (\xi_{SR})]~.
\label{xiSRxiLIV}
\end{equation}
We see here very vividly how the awkward mismatch between the superluminality of the electron
and classical Lorentz boosts amplifies the effects: in the correction term
the inevitable factor of $\delta$ is amplified by $\cosh^2 (\xi_{SR})$, which for a 91 GeV
electron is $\cosh^2 (\xi^{91GeV}_{SR}) \simeq 3.2 \times 10^{10}$. It should be clear
that this huge amplification sets the stage for when the boost transformation
from the lab frame to the rest frame must go
totally pathological, which is when the electron is actually superluminal (of course the bound
of Ref.~\cite{altschul} is derived assuming the electron would turn ``superluminal" only at some higher
energies, with the parameter $\delta$ small enough that at 91 GeV the electron still
actually has speed smaller than the speed-of-light scale).\\
The striking result is that one can use (\ref{xiSRxiLIV}), in light of (\ref{agreement}), to conclude
that $\delta  \lesssim 5 \times 10^{-15}$.

It should be clear that in the case of deformation rather than breakdown of Lorentz symmetry
the pathological amplification that produced this terrific bound on $\delta$ is not
to be expected. But in general even in a DSR framework there will be modifications
to the synchrotron radiated power: unlike the case of anomalous thresholds, a modification
of the synchrotron radiated power is not in conflict with the relativity of inertial
frames, so in general such a modification should be expected in a DSR framework.
And different DSR setups will produce different modifications of the synchrotron radiated power.
Consistently with the prudent attitude we are advocating we shall not attempt
to motivate some sort of general bound applicable to DSR senarios.
But we do consider explicitly two examples of such scenarios, just to show that
the line of analysis which Altschul correctly applied to the Lorentz-breaking case
fails to produce ``amplified bounds" when Lorentz symmetry is deformed.\\
In light of our purposes it is fitting to take as illustrative example the
case of deformed boost transformations which has been most studied
in the DSR literature.
This is the case of boost transformations such that (at leading order in $1/M_*$
and focusing for simplicity on the case of collinear boost and spatial momentum)
\begin{eqnarray}
&&\frac{dE}{d\xi} =   p ~,~~~
\label{boostA1}\\
&&\frac{dp}{d\xi} =  E - \frac{1}{M_*} E^2 - \frac{1}{2M_*} p^2
\label{boostA2}
\end{eqnarray}
which provide~\cite{dsr1,gacnewjourn,sethmajor,goldenrule}
 (also see, for a Hopf-algebra description of these non-linear laws,
 Refs.~\cite{majidruegg,kpoinap})
 a description compatible with the relativity of inertial frames
for the on-shell relation
\begin{equation}
m^2 = E^2 - {\bf p}^2 - \frac{1}{M_*} E {\bf p}^2
\label{metrictorsy}
\end{equation}
The reasons why this particular DSR setup has been of interest over the last decade
 are related to certain perspectives on the quantum-gravity problem,
which are here irrelevant. But is was important for our purposes to consider
 cases that have attracted interest independently of their possible implications
 for synchrotron radiation. This should be of simple reassurance to readers that
 no ``fine tuning of the DSR setup" can be seen as responsible for the result we shall
 now derive.

We just notice that
Eqs.~(\ref{boostA1}),(\ref{boostA2}) can be easily integrated to obtain the relationship
between the rest energy $m$ and the energy and momentum in a frame boosted with
respect to the rest frame by a rapidity $\xi$:
\begin{eqnarray}
&&E(\xi) = \cosh(\xi) m - \frac{1}{2 M_*} \sinh^2(\xi) m^2
\label{enexi}\\
&&p(\xi)  = \sinh(\xi) m - \frac{1}{2 M_*} \sinh(2\xi) m^2
\label{momxi}
\end{eqnarray}
Comparing with the corresponding special-relativistic formulas ($M_* \rightarrow \infty$ limit
of these formulas) one finds that the Altschul result (\ref{xiSRxiLIV}) for the broken-Lorentz
case is replaced in this DSR picture by
\begin{equation}
\cosh^4 (\xi) \simeq \cosh^4 (\xi_{SR}) \left( 1 + \frac{2m}{M_*} \cosh (\xi_{SR}) \right)
~.
\label{xiSRxiDSR1}
\end{equation}
Unsurprisingly this formula shows no peculiar amplification of the sort we highlighted
in relation to Eq.~(\ref{xiSRxiLIV}).
The correction term is just of the order of the superluminality this picture
endows to the electron, which is $E/M_* \simeq m \cosh (\xi)/M_*$.
This should be contrasted to the broken-Lorentz case, where the superluminality of the electron
was codified in $\delta$, and determinations of synchrotron radiated power
lead to $4 \delta \cosh^2 (\xi_{SR}) \leq 6 \times 10^{-4}$
({\it i.e.} $\delta  \lesssim 5 \times 10^{-15}$).
In our DSR picture, in light of (\ref{xiSRxiDSR1}), the line of analysis developed by
 Altschul only affords us a much weaker bound on
superluminality: $E/M_* \lesssim 3 \times 10^{-4}$.

While we are unable to establish general theorems on this issue (and such theorems
would anyway overshoot the objectives of our study), we do want to stress
that our observation applies well beyond the confines
of the case of DSR-compatible superluminality with linear dependence on energy.  For this purpose
we consider, borrowing from the Hopf-algebra literature~\cite{kpoinap,lukie1992},
a case where the superluminality is governed by $E^2/M_*^2$, in which the boosts
\begin{eqnarray}
&&\frac{dE}{d\xi} =  p  ~,
\label{boostB1}\\
&&\frac{dp}{d\xi} = E - \frac{1}{6M_*^2} E^3
\label{boostB2}
\end{eqnarray}
 provide a description compatible with the relativity of inertial frames
for the on-shell relation
\begin{equation}
m^2 = E^2 - {\bf p}^2 - \frac{1}{12 M_*^2} E^4 ~.
\label{metrictorsySB}
\end{equation}
From (\ref{boostB1}),(\ref{boostB2}) it follows that the rapidity $\xi$ that connects
a frame where the electron has energy $E$ to its rest frame is such that
\begin{equation}
E(\xi) = m  \cosh(\xi) - \frac{m^3}{48 M_*^2} \left[ \cosh^3(\xi) -\cosh(\xi)
+ 3 \xi \sinh (\xi) \right]
\nonumber
\end{equation}
For the large values of rapidity which are here of interest this formula can be compared
to its special-relativistic limit (of course obtained again for $M_* \rightarrow \infty$)
producing the estimate
\begin{equation}
\cosh^4 (\xi) \simeq \cosh^4 (\xi_{SR}) \left( 1 + \frac{m^2}{12 M_*^2} \cosh^2 (\xi_{SR}) \right)
~,
\label{xiSRxiDSRSTANDARD}
\end{equation}
Therefore, also this other DSR setup produced a prediction which is free from
the peculiar amplification we highlighted
in relation to Eq.~(\ref{xiSRxiLIV}).
The correction term is just of the order of the superluminality this picture
endows to the electron, which is $E^2/(8M_*^2) \simeq m^2 \cosh^2 (\xi)/(8M_*^2)$,
and on the basis of  (\ref{xiSRxiDSRSTANDARD}) the Altschul line of analysis
again produces only a weak bound: $E^2/(8 M_*^2) \lesssim 9 \times 10^{-4}$.

\section{Outlook on OPERA and indirect bounds}
The most frequently quoted laboratory bounds
on electron superluminality are indirect bounds derived assuming that
the laws of transformation among observers are undeformed
and there is a preferred frame. The strategy of analysis
adopted for deriving these bounds produces much weaker bounds
(even more than 10 orders of magnitude weaker) on
 some other pictures with electron superluminality, as we here showed
 considering the illustrative example of DSR deformations
of Lorentz symmetry.

Our main message would have been missed if the arguments here presented
were viewed as an attempt to establish the ``windows of opportunity"
for DSR deformations of Lorentz symmetry. As stressed in some of our previous
works~\cite{dsr1,whataboutopera},
from a theory perspective we cannot see any ``good reason" to expect
departures from Lorentz symmetry for LEP electrons any stronger than
at the level of a few parts in $10^{17}$. But in keeping scientific facts
cleanly separated from theory prejudice (even our own theory prejudice)
one cannot fail to be concerned for the possible implications of extensive
use of {\underline{indirect}} tests of the fundamental laws of physics. We gave here
an explicit example of this general concern: we believe that experiments aimed at
providing {\underline{direct}}
bounds on electron superluminality should
be in absolutely no
way discouraged on the basis of the availability of much ``tighter"
{\underline{indirect}} (and, as here shown, enormously model dependent) bounds.

We realize that the findings we here reported
may also invite speculations about a DSR interpretation
of the neutrino-superluminality anomaly tentatively 
reported by the OPERA collaboration~\cite{opera}.
If the data reported in Ref.~\cite{opera}
are taken at face value then some departures from Special Relativity
would have to intervene, since even the special-relativistic tachyon
is excluded~\cite{whataboutopera}.
And departures from Special Relativity that require a preferred frame are also
disfavored~\cite{cohenGla,operaDSR},
so one may want to try a DSR description~\cite{operaDSR,fransDSROPERA}.
Such speculations are legitimate, but they are somewhat premature not only
because the OPERA findings are still ``sub judice", but also 
because the DSR framework has not
 been yet developed comprehensively enough for such detailed phenomenological exercises.
It was somewhat fortunate for us that for the purposes of the thesis we were here putting forward
enough is understood of the DSR proposal to allow us to make a few valuable observations.
But the DSR research programme is still confronted by several ``open issues", including
 the fact that the structure of DSR-compatible
quantum field theories is still at a very early stage of development.
Moreover, one should take
 into account at least the following specific challenges:\\
(1) There are much stricter bounds
than the ones here discussed for the electron on DSR-deformations of Lorentz symmetry
for the photon, and they are {\underline{direct}} bounds.
For the DSR pictures we here considered these bounds exclude
anomalies up to a level of indeed a few parts in $10^{17}$ for
photons in the 1GeV-100GeV
range~\cite{grbgac,ellisFERMI,unoEdue,fermi090510}.\\
(2) The (subjective, but increasingly acknowledged) appeal of the DSR concept
has been grounded so far on the usefulness of this concept for the understanding
of certain quantum-spacetime pictures, and the corresponding quantum-spacetime
pictures have not provided so far much motivation for a particle-dependence of the
effects ({\it e.g.} wildly different effects for neutrinos and photons).\\
(3) The DSR setups which are so far better understood are of the type we here
considered, {\it i.e.} ``power-law energy-dependent deformations". But even taking the OPERA data
on neutrinos at face value, other available neutrino data appear to disfavor
a simple power-law energy dependence of
the effect~\cite{whataboutopera,giudiceOPERA,ellisOPERA,liberatiOPERA}.\\
(4) It should also be noticed that we took here the easy task of showing
that the arguments frequently cited to severely constrain superluminality of LEP electrons
are inapplicable to the DSR framework, but on the other hand we have not here
confronted the much more challenging task of identifying the best strategies for
constraining electron superluminality within a DSR picture. We expect that
a good candidate for establishing the best bounds on DSR-superluminality of the electron
should be the effects introduced by the recently uncovered relativity of spacetime
locality~\cite{bob,leeINERTIALlimit,prl}. It is emerging that just like in Special Relativity
the absoluteness of the speed scale $c$ comes at the ``cost" of the relativity of simultaneity,
most DSR frameworks with an absolute energy scale $M_*$ must pay the ``cost"
of a relativity of locality. Unfortunately the concept of relative locality
 is so new that it may take some
time to master it well enough for deriving robust bounds.

\bigskip

\bigskip

$~$

{\it We are happy to thank Paolo Lipari for encouragement and for very valuable
feedback on a preliminary version of this manuscript.}

$~$

$~$

\end{document}